\documentclass[%
 aip,
amsmath,amssymb,
preprint,%
]{revtex4-1}

\usepackage{graphicx}
\usepackage{dcolumn}
\usepackage{bm}
\usepackage[normalem]{ulem}

\usepackage[colorlinks,bookmarksopen,bookmarksnumbered,citecolor=red,urlcolor=red]{hyperref}

\usepackage{graphicx}
\usepackage{natbib}
\usepackage{color}
\usepackage{latexsym}
\usepackage{subfig}
\usepackage{natbib}
\usepackage{upgreek}
\usepackage{mathrsfs}
\usepackage{float}
\usepackage{caption}
\usepackage{soul}
\usepackage{textcomp}
\usepackage{stmaryrd}
\usepackage{amsmath}
\usepackage{lipsum}                     
\usepackage{xargs}                      
\usepackage[pdftex,dvipsnames]{xcolor}  
\usepackage[colorinlistoftodos,prependcaption,textsize=small]{todonotes}

\def\ii{{\rm i}}
\def\be{\begin{equation}}
\def\ee{\end{equation}}

\usepackage{verbatim}

\begin{document}

\preprint{AIP/123-QED}

\title[]{Wave interactions in neutrally stable shear layers: regular and singular modes, and non-modal growth}

\author{E. Heifetz}
\affiliation{Porter school of the Environment and Earth Sciences, Tel Aviv University,
 69978, Israel.}
\author{A. Guha}%
\affiliation{School of Science and Engineering, University of Dundee,  DD1 4HN, UK.}
\author{J.R. Carpenter}
 \affiliation{Institute of Coastal Research, Helmholtz-Zentrum Geesthacht, 21502, Germany.}


\date{\today}

\begin{abstract}

A recent letter [J. R. Carpenter and A. Guha,``Instability of a smooth shear layer through wave interactions'', Phys. Fluids, 31, 081701 (2019)] compared the neutral modes of a smooth two dimensional shear profile without an inflection point to the modes of its corresponding piecewise-linear profile.  The regular mode in the smooth profile was identified as the one least sensitive to the numerical resolution, while the singular modes displayed high sensitivity. 
Here we provide a physical interpretation using a wave interaction approach for understanding the structure and behavior of both the regular and singular modes. The regular modes are the interfacial Rossby waves located at the concentrated mean vorticity gradient of the shear profile. In contrast, the singular modes result from a one way phase-locking interaction between singular vorticity disturbances, passively advected by the mean flow at different levels of the profile, and the interfacial Rossby waves. We show that this one way interaction can also lead to a sustained non-modal growth of the interfacial Rossby waves that cannot be captured by standard eigenvalue analysis.

\end{abstract}

\maketitle

%

\section{Introduction}

The study of the stability of shear flows, pioneered by  Helmholtz \citep{professor1868xliii}, Kelvin \citep{thomson1871xlvi}, Rayleigh \citep{rayl1880}, Orr \citep{orr1907}, Sommerfeld \citep{sommerfeld1909}, Fj$\o$rtoft \citep{fjortoft1953}  and others,  has a rich history of a hundred and fifty years. Rayleigh formulated the eigen-problem for determining the linear stability of two-dimensional ($x$--$z$) inviscid shear flows to normal-mode perturbations\citep{drazin2004hydrodynamic,schmid2001stability} of the form $w(x,z,t) = \mathrm{Re}\{ \hat{w}(z)$ $\mathrm{e}^{\ii k(x-ct)} \}$, where $x$ and $z$ respectively denote the distance in the direction of the basic flow and the cross-stream direction, and $t$ represents time.  It is given by the celebrated Rayleigh equation 
\begin{equation}
    \label{eq_Rayl}
    \hat{w}_{,zz}-k^2\hat{w}- \dfrac{\overline{u}_{,zz}}{\overline{u}-c}\hat{w}=0;
    \quad z \in [-b,b],    
\end{equation}
where the comma subscript denotes derivative, $\overline{u}(z)$ is the basic flow profile in the $x$ (streamwise) direction, and $\hat{w}(z) \in \mathbb{C}$, $c \in \mathbb{C}$ and $k\in \mathbb{R}^+$ respectively denote the vertical (cross-stream) perturbation velocity eigenfunction, the corresponding streamwise phase speed and wavenumber.  A normal mode implies a vertical structure that grows/decays but does not deform, and travels with a constant speed.  When $\mathrm{Im}(c) \neq 0$, the mode grows exponentially, otherwise it is neutrally stable.  

Rayleigh's equation \eqref{eq_Rayl} is not a regular Sturm-Liouville eigenvalue problem \cite{bender2013advanced}.  One important consequence of this is that \emph{regular} normal modes, i.e., those that are free of singularities (hereafter called regular modes),  corresponding to the `discrete part' of the eigenspectrum, do not form a complete basis. Hence, when considering the initial value problem from which Eq. \eqref{eq_Rayl} is derived (i.e. Eq.~\eqref{eq:vorticity_eq}), an arbitrary initial disturbance cannot be represented by the superposition of discrete modes; it is necessary to include the `continuous spectrum' of \emph{singular} normal modes (hereafter, singular modes).  These singular modes arise from the singularity at the critical height, $z_s$, where the basic flow speed matches the modal phase speed: $\overline{u}(z_s) = c$, when the basic flow is stable and $c$ is real.  Simple inspection of Eq. \eqref{eq_Rayl} shows that $\hat{w}_{,zz}$ can be infinite at this location, implying a jump in the slope of the eigenfunction $\hat{w}(z)$. Aside from utilising a numerical solution, obtaining the singular mode eigenfunctions can be more involved than that of a discrete mode, and requires singular integral theory \citep{balmforth1995normal}. 

Considerably less attention has been given to the stable regular modes of Rayleigh's equation.  In fact, it has sometimes been stated that for stable normal modes the regular discrete spectrum is empty [see Drazin\citep{drazin2002introduction}, p.\, 148].  However,  
{ \citet{carp19} (hereafter, CG19)}
have demonstrated that there can be a regular mode that is embedded in the singular continuous spectrum in such flows.  This contradiction with Drazin\citep{drazin2002introduction} seems to have resulted from the fact that CG19 have finite regions where the background profile has $\overline{u}_{,zz} = 0$, that was likely not considered by Drazin\citep{drazin2002introduction}.  Note though, that even with a background profile where $\overline{u}_{,zz} \neq 0$, Iga\cite{iga2013} has suggested a method of recovering a regular discrete spectrum.  A primary goal of the present paper is to show that the identification of this regular mode allows for a physical interpretation of the singular continuous modes, revealing their structure, as well as properties of the spectra of stable flow profiles.  This is done by using a physical interpretation of the mechanisms of shear instability that has been referred to as wave interaction theory (WIT)\cite{hosk1985,bain1994,book,carp2013}.  

WIT arose because while normal mode solutions to the eigen-problem \eqref{eq_Rayl} are extremely useful in determining the stability of a given $\overline{u}(z)$ profile, it provides little insight into the physical mechanism(s) responsible for shear instabilities.  WIT is one tool that has been developed to address this lack of physical insight that often accompanies the mathematical solution of a normal mode analysis.  It began with a heuristic minimal model for shear instability based on the interaction at a distance between two counter-propagating Rossby waves (also referred to as vorticity waves) that occurs in shear layer profiles\cite{holm1962,bretherton,bain1994,heif2005,rede2001,carp2013,book}. This understanding {of WIT} has been formalized and extended over the last decades in various ways\cite{heif2019normal}, but continues to form the basis of WIT.  While WIT has been successful in physically describing modal and non-modal instabilities in terms of a phase-locked resonance of interfacial waves (Rossby, or otherwise), its application has been mainly limited to regular modes in piecewise-linear velocity profiles. In this paper we shall interpret the structure of singular continuous modes in smooth profiles.  Our analysis also serves as an extension of WIT to include singular neutral modes, and the utility of this extension is demonstrated by providing a physical interpretation of algebraic growth in shear flows.  We begin with a background into WIT in Sec. \ref{sec:WIT}, then describe a WIT-view of the regular and singular modes of a smooth, stable, shear flow that supports such wave motions.  In Sec.~\ref{sec:non-modal} we apply the knowledge gained in the previous section to form a WIT description of a non-modal instability arising from the interaction of modes of the singular continuous spectrum and the regular discrete mode.  The final section of the paper discusses and summarises the main results. 


\section{Wave interaction theory (WIT)} \label{sec:WIT}

Our starting point is the two-dimensional ($x$--$z$) inviscid, linearized vorticity equation around a general mean state defined by the background profile $\overline{u}(z)$, and $\overline{q}_{,z}= -\overline{u}_{,zz} $, i.e.,
\begin{equation} \label{eq:vorticity_eq}
q'_{,t}+\overline{u} q'_{,x}=-w'\overline{q}_{,z},
\end{equation}
and quantities with primes {denote} perturbation quantities. The perturbation velocity field is ${\bf u}' = (u',w')$
and $q' = w'_{,x}-u'_{,z}$  is the vorticity perturbation.  The flow is taken to be bounded in the vertical direction by impenetrable boundaries at $z=\pm b$. 
We look for wavelike solutions of the form 
\begin{equation} \label{eq:vorticity_k}
q'_k = \textrm{Re}\{ \hat{q}(z,t;k)\mathrm{e}^{\ii kx}\} = \textrm{Re}\{Q(z,t;k)\mathrm{e}^{\ii[kx+\epsilon(z,t;k)]}\},
\end{equation}
where we have explicitly split the complex vorticity perturbation, $\hat{q}$, into amplitude, $Q(z,t;k)$, and phase, $\epsilon(z,t;k)$, components, which are both real.  Hereafter we treat each Fourier component with wavenumber $k$ separately and for convenience drop both the $k$ subscript and the prime superscript. 

{
Since the flow is incompressible we may define all
variables in terms of the streamfunction}
$\psi=\textrm{Re}\{ \hat{\psi}(z,t,k)\mathrm{e}^{\ii kx}\}$ as follows:
$u = -\psi_{,z};\, w= \psi_{,x} = \ii k\psi;\, q=\nabla^2\psi = -k^2\psi + \psi_{,zz}$.  Given these definitions it is possible to write the cross-stream velocity in terms of the vorticity so that
\begin{equation} \label{eq:psi}
\hat{w} = - \ii k\int_{-b}^{b} \hat{q}(\tilde{z},t,k)G(z,\tilde{z})d\tilde{z},
\end{equation}
with the (positive definite) { Green's} function \cite{heif2005}
\begin{equation} \label{eq:Green}
G(z,\tilde{z}) =  {1\over k \sinh{(2kb)}}\Bigg\{ \begin{array}{l@{\quad}l}
\sinh{[k(b+\tilde{z})]}\sinh{[k(b-z)]}, & \textrm{$\tilde{z} \leq z \leq b$} \\ 
\sinh{[k(b-\tilde{z})]}\sinh{[k(b+z)]}, & \textrm{ $-b \leq z \leq \tilde{z}$},\\ \end{array} 
\end{equation}
satisfying { $ G_{,zz}-k^2G = - \delta(z-\tilde{z})$} and the boundary conditions $G(z= \pm b,\tilde{z})=0$ for zero normal velocity at the solid boundaries  $z=\pm b$. Equation \eqref{eq:psi} shows that the  vertical velocity (as well as the streamfunction since they are related via $\hat{w} = \ii k\hat{\psi}$)
at any location $z$ can be expressed in terms of the influence of vorticity perturbations throughout the domain. 
After using Eq.~(\ref{eq:vorticity_k}) and Eq.~(\ref{eq:psi}) in  Eq.~(\ref{eq:vorticity_eq}), and performing some algebra, we obtain
\begin{equation} \label{eq:complex_q}
{Q_{,t}(z,t)\over Q(z,t)} + \ii\epsilon_{,t}(z,t)=  \ii k\left[-\overline{u}(z) + \overline{q}_{,z} (z)
\int_{-b}^{b}{Q(\tilde{z},t)\over Q(z,t)}\mathrm{e}^{-\ii\Delta\epsilon(z,\tilde{z},t)}
G(z,\tilde{z})d\tilde{z}\right],
\end{equation}
where 
we define the phase difference $\Delta\epsilon(z,\tilde{z},t)\equiv \epsilon(z,t) - \epsilon(\tilde{z},t)$. Taking the real and imaginary parts of  Eq.\,(\ref{eq:complex_q}) leads to
\begin{subequations}
\begin{align}
 \label{instant_gr}
\mathrm{{growth} \: rate}& = {Q_{,t}(z,t)\over Q(z,t)} = k\left[\overline{q}_{,z} (z)
\int_{-b}^{b}{Q(\tilde{z},t)\over Q(z,t)}\sin\left({\Delta\epsilon(z,\tilde{z},t)}\right)
G(z,\tilde{z})d\tilde{z}\right], \\
\label{instant_ps}
\mathrm{{phase} \: speed} &= -{\epsilon_{,t}(z,t)\over k} = \overline{u}(z) -
\overline{q}_{,z} (z)
\int_{-b}^{b}{Q(\tilde{z},t)\over Q(z,t)}\cos\left({\Delta\epsilon(z,\tilde{z},t)}\right)
G(z,\tilde{z})d\tilde{z}\, .
\end{align}
\end{subequations}
{Equations (\ref{instant_gr})--(\ref{instant_ps})} describes the local instantaneous growth rate and phase speed of a vorticity perturbation at level $z$ arising from the net action of  $w$ velocities over the entire domain. 
For the case of normal mode solutions, where $q(x,z,t) = \mathrm{Re}\{ \hat{q}(z) e^{\ii k(x-ct)} \}$ with $c$ constant, the term inside the square brackets on the RHS of Eq. (\ref{instant_gr}) is the imaginary part of the phase speed, $c_i$, and the entire RHS of Eq. (\ref{instant_ps}) is the real part of the phase speed, $c_r$. Hence, for normal modes, both of the RHSs are constants independent of $(z,t)$.  Furthermore, for neutral modes, the RHS of Eq. (\ref{instant_gr}) is zero since $c_i = 0$. 

\begin{figure*}
\centering\includegraphics[width=0.8\textwidth]{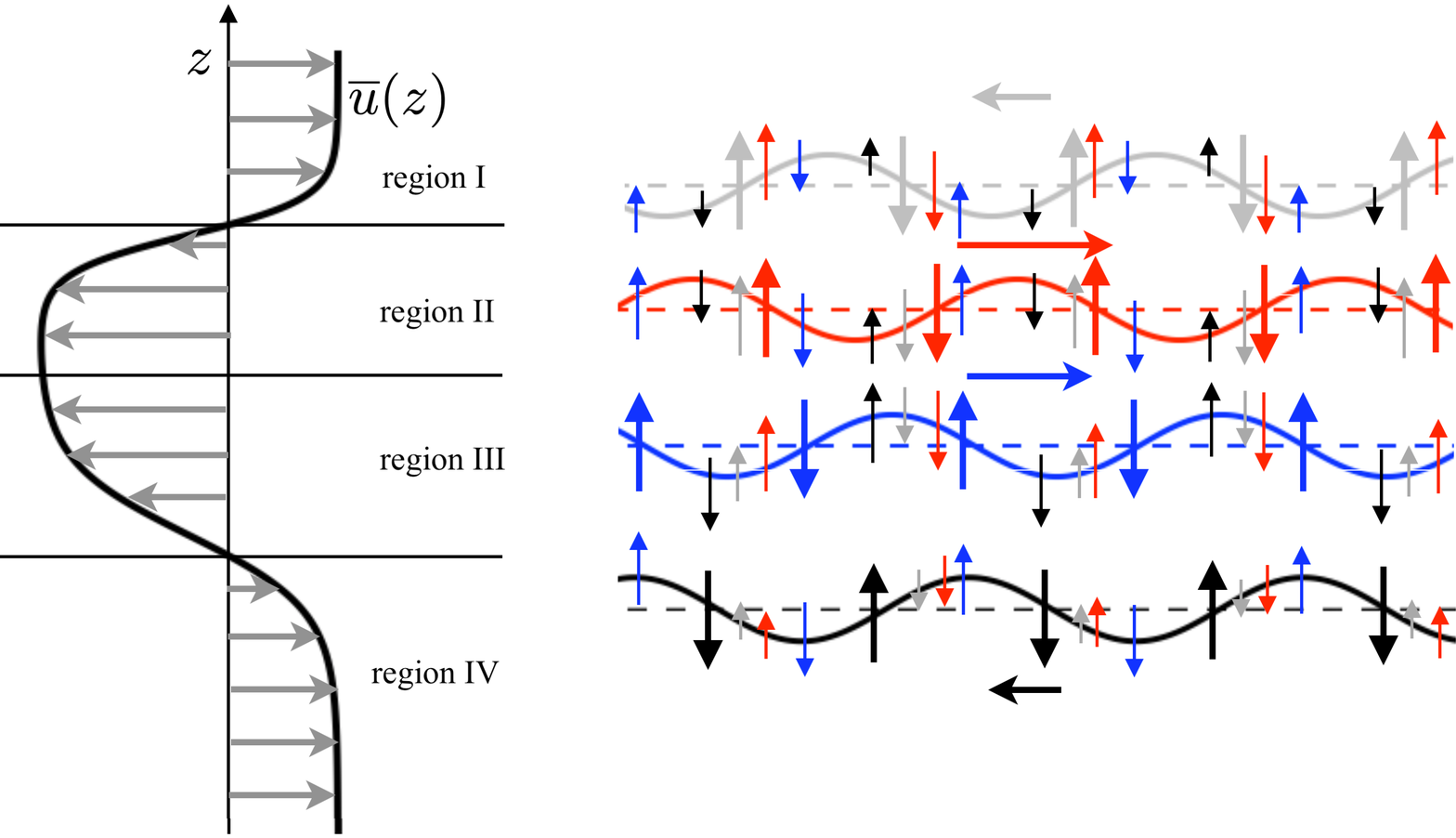}
\caption{(Color online) Schematic of a general shear layer instability, which can be understood 
in terms of the action-at-a-distance interaction between Rossby/vorticity waves (here illustrated by 4-wave interaction). 
The cross-stream velocity at each layer is
attributable to the vorticity field throughout the whole shear layer, induced by all waves, and attenuated according to the Green function $G(z,\tilde z)$ (indicated by the attenuated vertical arrows of the same color). According to the phase difference between the waves, and the sign of $\overline{q}_{,z}$, the waves affect each other's propagation speed and amplitude growth. The waves are drawn centred on regions of local extrema of $\overline{q}_{,z}$.}
\label{fig:Multi_waves}
\end{figure*}

The action-at-a-distance mechanistic interpretation of WIT, expressed in
Eqs. (\ref{instant_gr})--(\ref{instant_ps}) is demonstrated schematically in Fig. \ref{fig:Multi_waves} (for details see \citet{heif2019normal}).
{The vorticity at each level changes its amplitude and
phase due to the vertical advection of the mean vorticity gradient, $\overline{q}_{,z}$, by the cross-stream velocity $w$. The latter is 
attributable to the vorticity perturbation field throughout the whole domain, and attenuated according to the Green function $G(z,\tilde z)$.}
When two remote vorticity perturbations are in perfect quadrature, {i.e.}\,
$\sin\left({\Delta\epsilon(z,\tilde{z},t)}\right)=\pm 1$, the interaction affects only the amplitude, but when those waves are perfectly in phase or anti-phase, {i.e.}\,
$\cos\left({\Delta\epsilon(z,\tilde{z},t)}\right)=\pm1$, 
they affect each other's propagation rate in the streamwise direction with no change in the amplitude. 

\section{Understanding regular and singular modes from a WIT perspective}

\subsection{Basic flow}

\begin{figure}
\begin{center}
\includegraphics[width=0.9\textwidth]{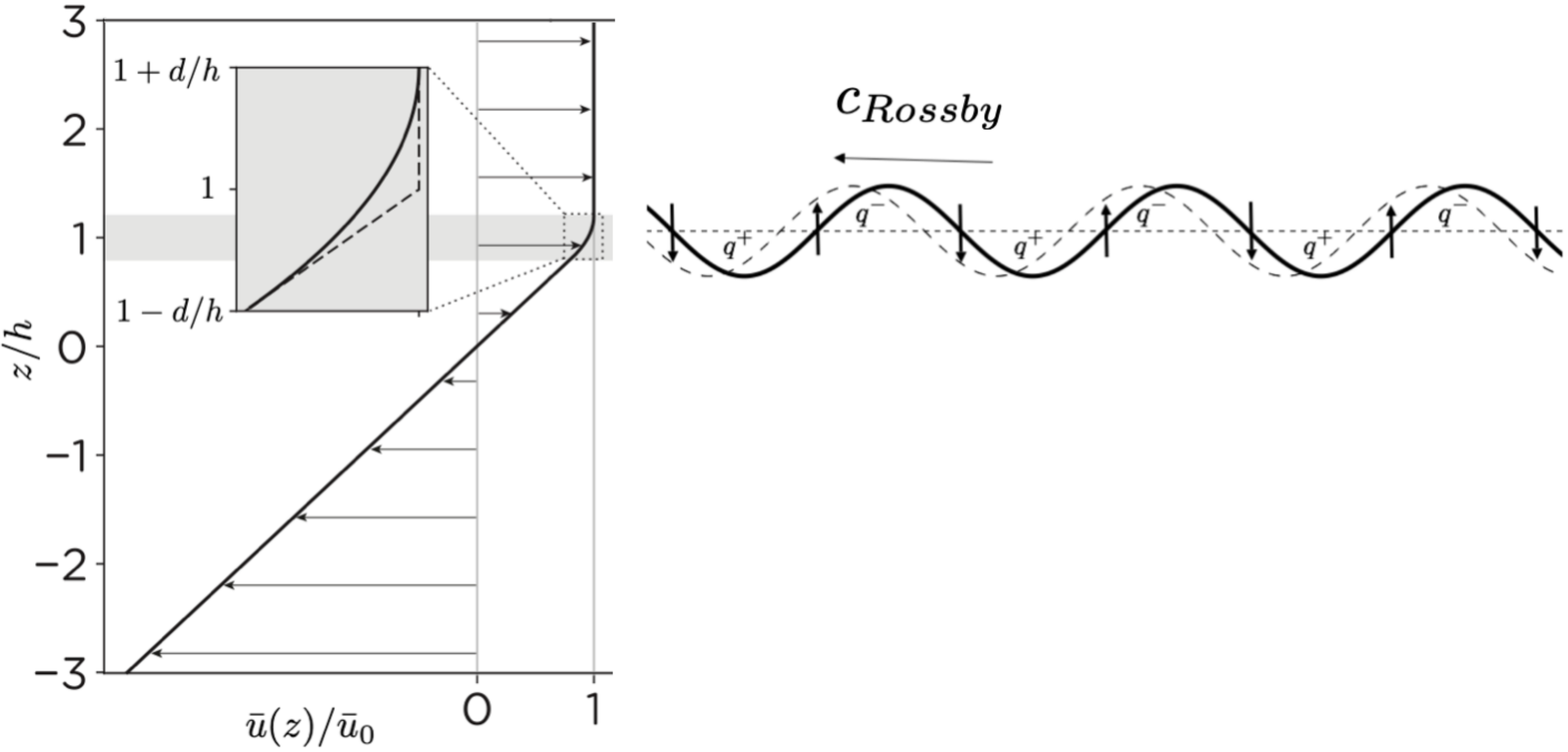} 
\caption{Profiles of a smooth vorticity interface exhibiting stable vorticity waves.
Region in which $\overline{q}_{,z} \neq 0$ is highlighted in grey, and has  a thickness of $d = 0.2h$, with vertical boundaries located at $z=\pm b=\pm 3h$.  Piecewise-linear profile differs from the smooth profile only in the grey region, and is shown as an inset. The right hand side shows an interfacial Rossby wave that would exist in the piecewise-linear profile. The variables {$q^+$ and $q^-$ respectively denote  the counterclockwise and clockwise vorticity wave anomalies.} The vertical arrows represent $w$ velocity, bold  curve represents current position, while dashed one represents the position after a short time interval.  }
\label{f:profiles}
\end{center}
\end{figure}

We now wish to provide a mechanistic interpretation using WIT of both regular and singular modes occurring in smooth and piecewise-linear profiles.  To fix ideas, we consider the smooth shear profile shown in Fig.~\ref{f:profiles}, previously suggested by Baines et al.\cite{bain1996}, {defined by}
\begin{equation} \label{eq:baines_profile}
\overline{u}(z)
= \overline{u}_0 \left\{ \begin{array}{l@{\quad}l} 1  ,& \textrm{ $b \geq z\ge h + d$} \\
1 - (h + d - z)^2/(4 h d) ,& \textrm{ $|z-h| \le d$} \quad \\ 
z/h ,& \textrm{ $-b \leq z \le h - d$} .\\ \end{array}  \right.
\end{equation}
This {flow} profile consists of a finite region of non-zero, but constant, background vorticity gradient, $\overline{q}_{,z} = \overline{u}_0/2hd$, in the region $|z-h|<d$, surrounded by constant sheared regions where $\overline{q}_{,z} = 0$.  Since {this basic flow profile} lacks an inflection point, it is stable to normal mode perturbations, and would not be expected to yield any regular modes (e.g., Drazin\citep{drazin2002introduction}, p.~148). However, its limiting case when $d/h\rightarrow 0$ is the piecewise-linear profile
\begin{equation} \label{eq:piecewise}
\overline{u}(z)
= \overline{u}_0 \Bigg\{ \begin{array}{l@{\quad}l} 1  ,& \textrm{ $b \geq z \ge h$} \\ 
z/h ,& \textrm{ $-b \leq z \le h $},\\ \end{array} 
\end{equation}
which has a regular discrete mode with a phase speed given by
\begin{equation} \label{eq:regular_c}
c_{regular} = \overline{u}_0\left\{1-
{\sinh{[k(b+h)]}\sinh{[k(b-h)]}\over kh \sinh{(2kb)}}\right\} = 
\overline{u}(h) - c_{Rossby} .
\end{equation}
Here $c_{Rossby}$ is the intrinsic interfacial Rossby wave phase speed (defined positive here) in a bounded domain.  

If the kink in $\overline{u}(z)$ at $z=h$ is smoothed by using the profile in Eq. \eqref{eq:baines_profile}, CG19 showed numerically that a corresponding regular mode does exist.  It is embedded within the singular continuous spectrum, and is not immediately recognisable as a regular mode. This mode is indeed regular because unlike singular modes, $\hat{w}$ is differentiable at the critical height. Furthermore, when $d/h \rightarrow 0$, the phase speed of this mode converges to Eq. \eqref{eq:regular_c}.  It can be concluded that this regular mode is the smooth analogue of the interfacial Rossby wave from the piecewise-linear profile. 


\subsection{Regular mode}
This stable regular mode that is present in the piecewise and smooth profiles is associated with an interfacial Rossby wave (also called a vorticity wave).  It propagates on changes in the background vorticity $\overline{q}_{,z}$, much like classical Rossby waves  {propagating} on changes in  {the} planetary vorticity.  It therefore results solely from the vorticity dynamics in the grey area of 
Fig. \ref{f:profiles}. The propagation of the wave occurs through the vertical displacement of the vorticity gradient inducing a vertical velocity that is $\pi/2$ radians out of phase with the displacement field, shown in Fig.~\ref{f:profiles}.  



The regular mode for the piecewise-linear profile given by Eq. \eqref{eq:piecewise} has an infinite vorticity perturbation at the level of the kink, due to the vertical displacement of an infinite gradient, i.e., $q = -\overline{q}_{,z} \zeta$, with $\zeta = w/[\ii k(\overline{u} - c)]$ the vertical displacement.  It can be expressed as
\renewcommand{\theequation}{\arabic{equation}}
\begin{equation} \label{q_regular1}
q_{regular}= \textrm{Re}\{\hat{q}\delta(z-h)\mathrm{e}^{\ii kx}\}= \textrm{Re}\{\underbrace{\hat{Q}(h,t)\delta(z-h)}_{Q(z,t)} \mathrm{e}^{\ii[kx + \epsilon(h,t)]}\,\} .
\end{equation}
The Eqs.\,(\ref{instant_gr})--(\ref{instant_ps}) are also applicable for a single $\delta$-function vorticity wave, implying $\Delta\epsilon(z,\tilde{z}) = \Delta\epsilon(h,\tilde{z}) = \Delta\epsilon(h,h) =0$, so that the RHS of Eq.\,(\ref{instant_gr}) vanishes. This also implies that $Q$ is independent of $t$.
Substituting Eq.\,(\ref{eq:piecewise}) and Eq.\,(\ref{q_regular1}) in Eq.\,(\ref{instant_ps}) for $z=h$, and recalling that $\overline{q}_{,z} = ({\overline{u}_0}/h)\delta(z-h)$, we obtain
\begin{align}
c_{regular}  & = \overline{u}_0\left[1-{\delta(z-h)\over h}\int_{-b}^{b}{\hat{Q}(\tilde{z})\delta(\tilde{z}-h)\over \hat{Q}(h)\delta(z-h)}G(h,\tilde{z})d\tilde{z}\right]
\nonumber \\ 
& = \overline{u}_0\left[1-{1\over h\hat{Q}(h)}\int_{-b}^{b}\hat{Q}(\tilde{z})\delta(\tilde{z}-h)G(h,\tilde{z})d\tilde{z}\right] 
 = 
 \overline{u}_0\left[1- {G(h,h)\over h}\right],
\label{c_regular2}
\end{align}
which is identical to Eq.\,(\ref{eq:regular_c}) when $G(h,h)$ is substituted into Eq.\,(\ref{eq:Green}), and 
\begin{equation} \label{C_Rossby}
c_{Rossby} = \overline{u}_0 G(h,h)/ h.
\end{equation}
We note here that while the wave's vorticity field is concentrated at $z=h$, its induced streamfunction, and hence its induced velocity field, fill the entire domain. This can be readily verified by substituting Eq. (\ref{eq:Green}) and Eq. (\ref{q_regular1}) in Eq. (\ref{eq:psi}):
{
\begin{equation} \label{psi_regular}
\psi_{regular}(x,z,t)= \mathrm{Re}[ - \hat{Q}(h,t)G(z,h) \mathrm{e}^{\ii\{kx + \epsilon(h,t)\}}]\, .
\end{equation}}
Due to the structure of the { Green's} function, $w=\psi_{,x}$ is continuous everywhere but $u = -\psi_{,z}$ is discontinuous at  $z=h$. The latter yields an infinite shear perturbation $-u_{,z}$ at $z=h$, which accounts
for the $\delta$-function structure of the vorticity perturbation there. 

\begin{figure}
\centering\includegraphics[width=0.7\textwidth]{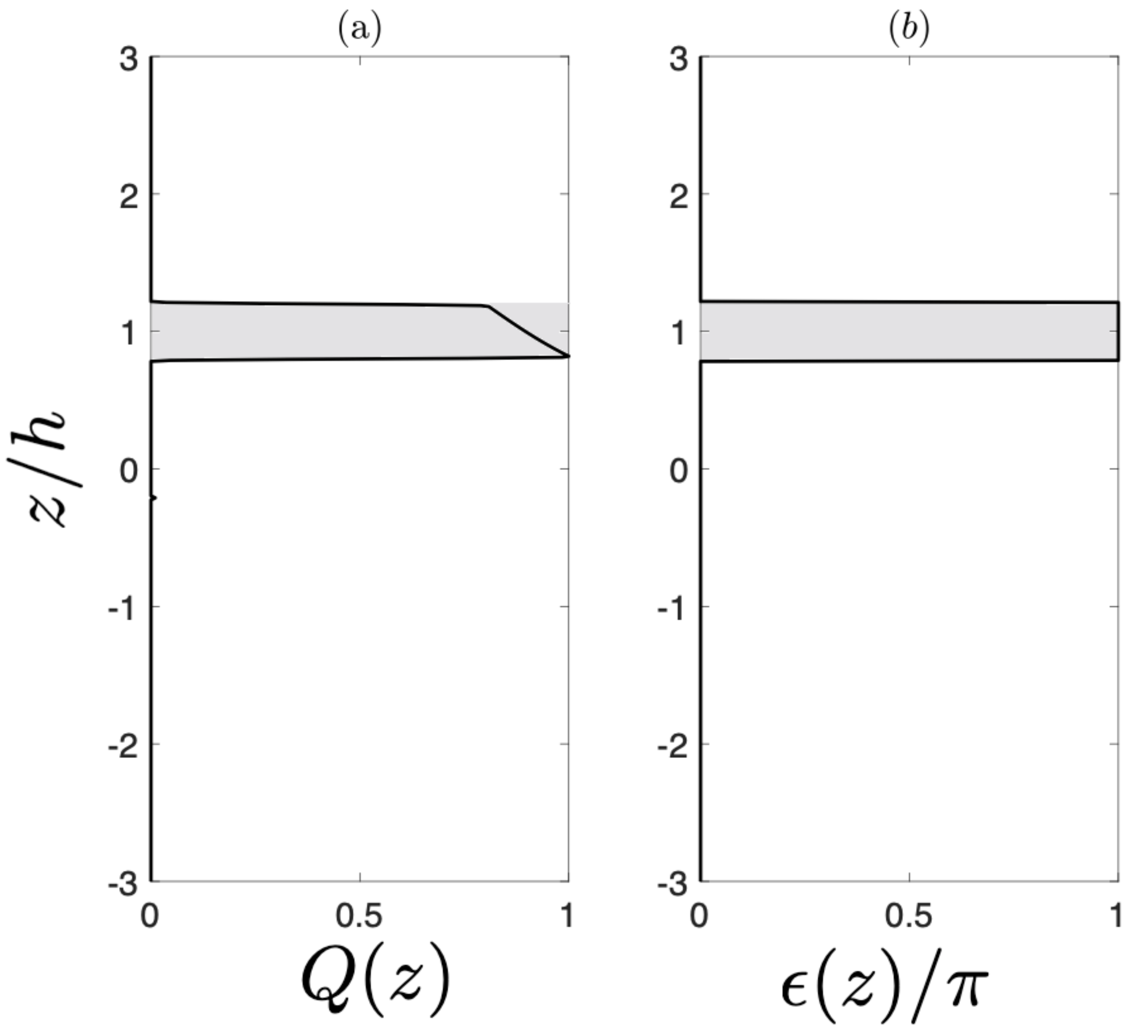}
\caption{Vorticity perturbation eigenfunction of the regular mode of the smooth profile in Eq.\,(\ref{eq:baines_profile}).  This profile has non-zero vorticity gradient ($\overline{q}_{,z} \neq 0$) in the grey band centred on $z/h = 1$. (a) Amplitude $Q(z)$  and (b) phase $\epsilon(z)/\pi$.
}
\label{fig:eigen}
\end{figure}

For smooth profiles, that we represent with Eq.\,(\ref{eq:baines_profile}), the regular mode structure (amplitude and phase) has been obtained numerically following CG19, and is plotted in Fig.\,\ref{fig:eigen}. The vorticity perturbation is non-zero only in the region $|z-h|<d$ where $\overline{q}_{,z}\neq 0$ (grey region of Fig.\,\ref{f:profiles},\ref{fig:eigen}), and has constant phase with an amplitude $Q$ that decays monotonically with $z$ inside the grey region. The neutrality of the mode in Eq.\,(\ref{instant_gr}) is assured since the vorticity perturbations at every $z$ level are in phase with $\Delta\epsilon(z,\tilde{z},t)=0$ for every pair of $(z,\tilde{z})$ in the grey region $|z-h| < d$.  Consequently, the vertical (cross-stream) velocity field that each level of the vorticity perturbation induces on each other are in phase.  This causes the entire perturbation to propagate counter to the mean flow as a vorticity/Rossby wave. Since all the perturbations within the grey zone propagate in concert, with the same normal mode phase speed $c_{regular}$, and since the mean flow $\overline{u}(z)$ increases with height, the perturbations at the upper levels of the grey zone require more help to counter-propagate against the mean flow than the waves in the lower levels. This is achieved by the monotonic decrease of $Q$ with height, as seen in Fig.\,\ref{fig:eigen}(a). Since the far-field velocity induced by each wave is proportional to the magnitude of its vorticity, the lower perturbations help the upper perturbations (in the grey zone) to counter-propagate more effectively than vice versa.


This argument can be stated mathematically by substituting Eq.\,(\ref{eq:baines_profile}) and $\Delta\epsilon(z,\tilde{z},t)=0$ in Eq.\,(\ref{instant_ps})
for $|z-h|<d$, to obtain the following transcendental equation for $Q(z)$: 
\begin{equation} 
c_{regular} =
\overline{u}_0\left[1-{(h+d-z)^2\over 4hd}-
{1\over 2hd}\int_{h-d}^{h+d}{Q(\tilde{z})\over Q(z)}
G(z,\tilde{z})d\tilde{z}\right]\, .
\end{equation}
whose solution agrees with that obtained by CG19 by a different method.  As $d/h \to 0$, $c_{regular}$ converges to the solution of Eq.\,(\ref{c_regular2}), however, even for $d/h = 0.2$ the agreement is excellent (as shown by CG19 in their Fig.\,2e).

\subsection{Singular modes}


Regular modes are on the same isovortical manifold of the basic state, meaning that they are obtained from (sinusoidal) deformations of the horizontal vorticity material lines of the basic state. Since $w ={D\zeta/ Dt}$,
 under linearization it yields $w =  \zeta_{,t}+\overline{u}\zeta_{,x}$. Substituting back {into} Eq.\,(\ref{eq:vorticity_eq}) we obtain that isovortical perturbations satisfy
\begin{equation} \label{isovortic}
q^{(IS)} = \underbrace{-\overline{q}_{,z}\zeta}_{\textrm{isovortical deformation}},
\end{equation}
i.e., isovortical vorticity perturbations are given through a simple vertical displacement of the background vorticity gradient.

In regions where $\overline{q}_{,z}=0$, any non-zero vorticity perturbation must be non-isovortical, $q^{(NI)}$, that was seeded as an initial condition. These non-isovortical vorticity perturbations are passively advected by the mean flow at that location,
\begin{equation} \label{non_isovortic}
q_{,t}^{(NI)}+\overline{u}{q}_{,x}^{(NI)}=0.
\end{equation}
If at  $z=z_s$ we have $\overline{q}_{,z}=0$, then a solution for Eq.\,(\ref{non_isovortic}) can have the form of 
{
\begin{equation} \label{delta_q_s}
q^{(NI)} = \mathrm{Re}[ \hat{q} \delta(z-z_s)\mathrm{e}^{\ii kx}] =\mathrm{Re}[ \hat{Q}_s\delta(z-z_s)
\mathrm{e}^{\ii\left\{kx+\epsilon(z_s,t)\right\}}],
\end{equation}}
where $\hat{Q}_s\equiv \hat{Q}(z_s)= \mathrm{constant}$,  the phase $\epsilon(z_s,t) =  \epsilon_0(z_s) - k\overline{u}(z_s)t\,$, and $\epsilon_0$ is the phase at $t=0$.


Here we show that the singular modes, observed by CG19, are obtained by the interaction between vorticity pairs in the form of Eqs.\,(\ref{isovortic}) and (\ref{delta_q_s}). Since the far-field velocity induced by the regular modes of the smooth and the piecewise-linear profiles are almost identical (provided that $d/h$ is small enough), we will hereafter consider the piecewise-linear profile to investigate the singular modes, almost without any loss of accuracy. Therefore, the singular modes can  be written as
\begin{equation} \label{singular_q}
q_{singular} =  \hat{q}(h)\delta(z-h) + \hat{q}(z_s)\delta(z-z_s),
\end{equation}
for any level of $z_s$ in the domain that differs from $z=h$, i.e., satisfying $z_s \in (-b,b)\backslash\{h\}$.
We denote $\hat{q}(z_s)$ as a ``passive" vorticity wave as it is simply advected by its local mean flow, and undergoes no change in amplitude. In contrast, we denote $\hat{q}(h)$ as an ``active" vorticity (Rossby) wave as it acts to propagate counter to its local mean flow.  It may change its amplitude through changes in vertical displacement [i.e., Eq.~(\ref{isovortic})] caused by interaction at a distance with the passive vorticity wave.  Since $\hat{q}(z_s)$
must be advected by the local mean flow at $z=z_s$, the phase speed of the singular mode must be $c_{singular}(z_s) = \overline{u}(z_s)$, as all properties of the modal solutions are moving in concert.
Furthermore, since the singular modes are neutral,  the two vorticity perturbations must be either in phase or anti-phased, i.e., $\Delta\epsilon(h,z_s) = 0$ or $\pi$. The ``active'' vorticity (Rossby) wave at $z=h$ cannot affect the phase speed of the 
``passive'' wave at $z=z_s$, but the passive wave affects the propagation of the active one by inducing the cross-stream velocity far field at $z=h$. When the waves are in phase, the passive helps the active to advect the mean vorticity in the cross-stream direction, and thus makes the active wave propagate faster against the local mean flow. Conversely, when they are anti-phased, the passive wave hinders the counter-propagation rate of the active one. Therefore, singular modes are the consequence of a phase locking that tunes the active wave to propagate with the phase speed of $\overline{u}(z_s)$. Hence, if 
$c_{regular} > \overline{u}(z_s)$, the waves are in phase, whereas if $c_{regular} < \overline{u}(z_s)$, they are {in} anti-phase.

This can be shown mathematically when we substitute the singular mode solutions at $z=h$ in  Eq.\,(\ref{instant_ps}) and use Eq.\,(\ref{eq:regular_c}) and Eq.\,(\ref{C_Rossby})  to obtain
{
\begin{equation} \label{Cs}
c_{singular}= \overline{u}(z_s) = \overline{u}_0 - \left[ c_{Rossby} \pm{\overline{u}_0 \over h}
{\hat{Q}(z_s)\over\hat{Q}(h)}G(h,z_s)\right] = c_{regular}\mp{\overline{u}_0 \over h}
{\hat{Q}(z_s)\over\hat{Q}(h)}G(h,z_s),
\end{equation}}
where the upper sign in the ($\pm$, $\mp$) accounts for the in-phase solution and the lower to the anti-phased one.  The singular mode phase speed is therefore the result of alterations to the regular mode (first term) caused by the interaction with the singular passive wave (second term).

It will be as well helpful to define the steering level of the regular mode, $z_{sl}$, as the level in which $\overline{u}(z_{sl}) = c_{regular}$. Here we emphasize the difference between the better known critical level with the steering level;   while the former is defined as the location $z=z_c$ where $\overline{u}(z_{c}) = c$ and may have $\overline{q}_{,z}(z_{c})\neq 0$, the latter is the location $z=z_{sl}$ where $\overline{u}(z_{sl}) = c_{regular}$ with $\overline{q}_{,z}(z_{sl})=0$. Note that only vanishing background vorticity gradients at the critical and steering levels are considered in our present set-up. From Eqs. (\ref{eq:piecewise}) and  (\ref{c_regular2}) we obtain that $z_{sl} =h-G(h,h)$.

\begin{figure}
\centering
\includegraphics[width=0.9\linewidth,keepaspectratio]{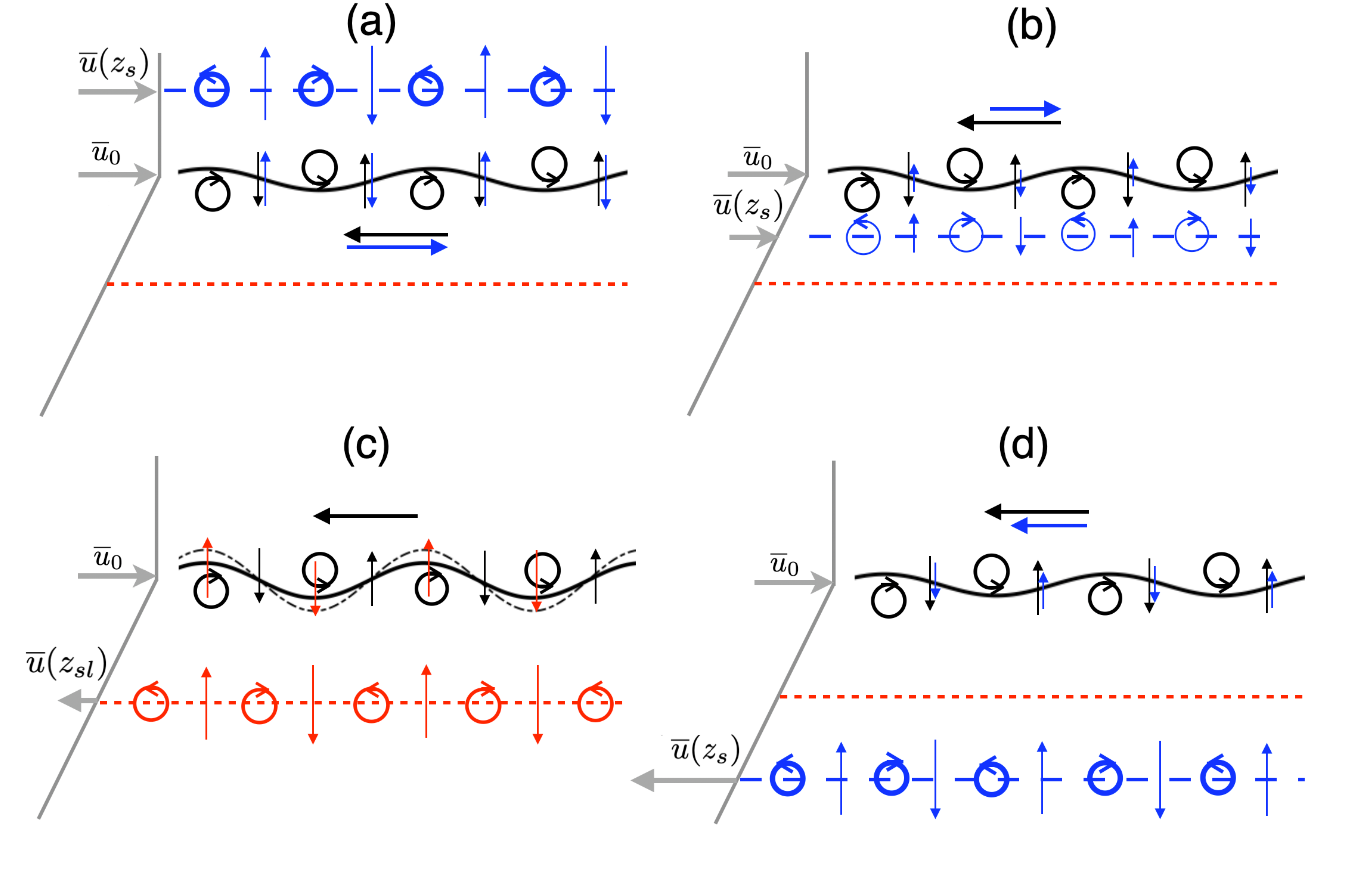}
\caption{ { Depending on the location of the passive wave, there could be four cases: (a) Passive wave located above the active wave, (b) passive wave located below the active wave but above the steering level, (c) passive wave located at the steering level, and (d) passive wave located below the steering level. Grey color denotes the background velocity, black wavy line  denotes the active wave's displacement field, blue dashed line denotes the passive wave, while red dashed line represents the steering level. In (c), the passive wave is colored red since it coincides with the steering level, resulting in a non-modal growth of the active wave (growth represented by dashed black line). In all cases, vertical arrows denote $w$, the cross-stream velocity, and circles with arrows denote the sign of the vorticity perturbation (higher strength denoted by thicker circles). The black horizontal arrows represent intrinsic phase speed of the active wave, while blue horizontal arrows denote the phase speed induced by the passive wave on the active wave.} }
\label{fig:4cases}
\end{figure} 

\begin{figure}
\centering
\includegraphics[width=0.7\linewidth,keepaspectratio]{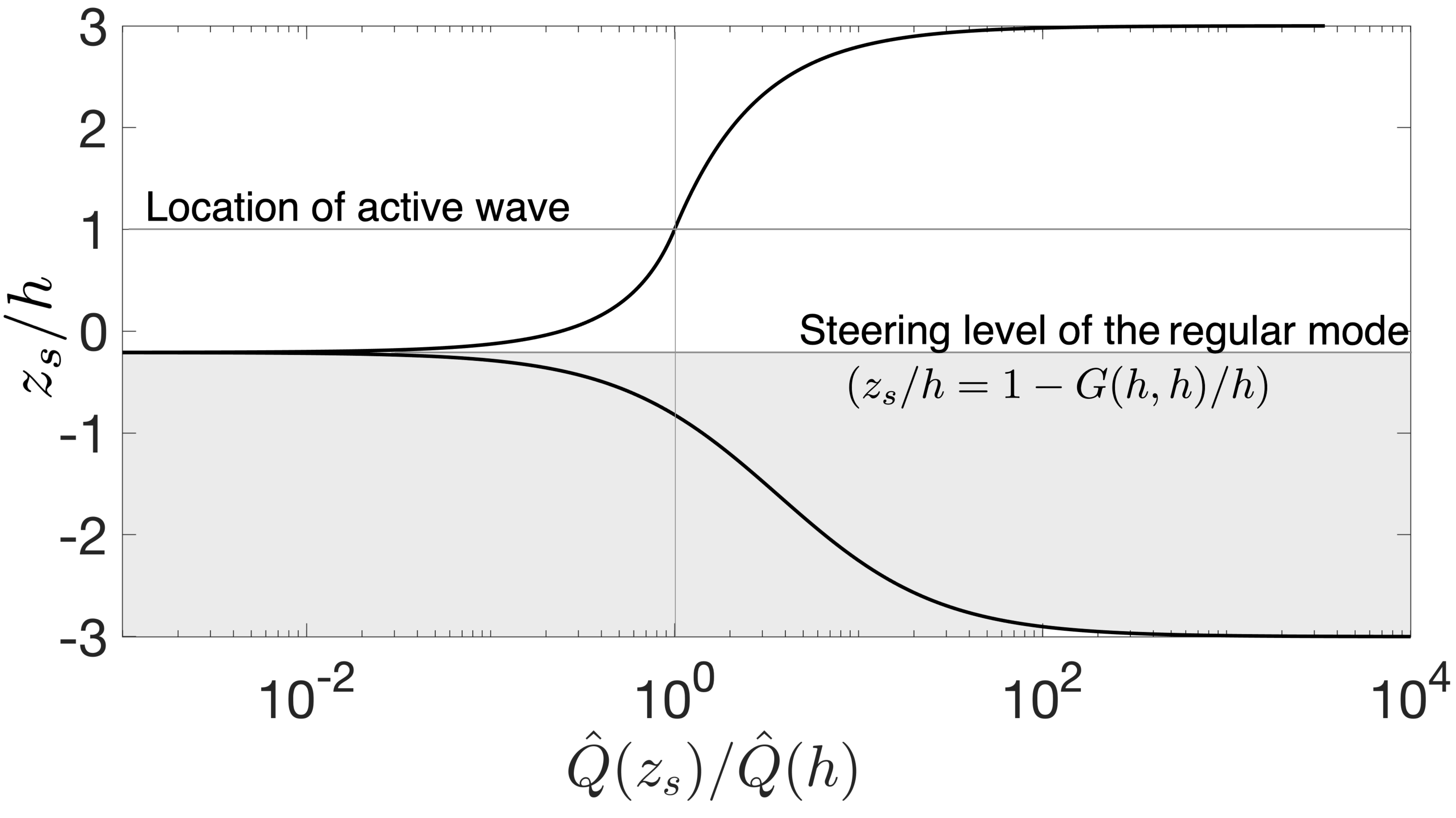}
\caption{Vorticity amplitude ratio of the passive to the active waves, $\hat{Q}(z_s)/\hat{Q}(h)$, as a function of the passive wave position, $z_s$. Active wave is located at $z_s/h=1$, where the amplitude ratio is unity.
The location where the amplitude ratio vanishes ($z_s/h=1-G(h,h)/h$) is the steering level of the regular mode.  
For $z_s$ below this level the passive and active waves are in phase (grey region), and above they are in anti-phase (white region).}
\label{fig:eigen_ratio}
\end{figure}

We can now identify three cases that set the general structure of the singular modes:

   
\begin{enumerate}
\item  \emph{Passive wave located above the active wave} $[b\!>\!z_s\!>\!h\, \Longleftrightarrow c_{singular} = \overline{u}(z_s) = \overline{u}_0]$: In this case the waves are anti-phased with the amplitude ratio ${\hat{Q}(z_s)/\hat{Q}(h)} = {G(h,h)/ G(h,z_s)}>1$, so that the passive wave has a greater amplitude of vorticity perturbation than the active wave. The cross-stream velocity field induced by 
$\hat{Q}(z_s)$ at $z=h$ via Eq.~(\ref{eq:psi}) is strong enough to completely nullify the intrinsic propagation speed, $c_{Rossby}$, caused by the self-induced cross-stream velocity due to $\hat{Q}(h)$ at $z=h$. Hence although $\overline{q}_{,z}$ is non zero at $z=h$, the cross-stream velocity vanishes there and consequently the vorticity perturbation is simply advected by the mean flow. This set-up is shown in Fig. \ref{fig:4cases}(a).

\item  \emph{Passive wave is located below the active, but above the regular mode steering level $z_{sl}$} 
[$z_{sl}  = h-G(h,h)\! <\! z_s \!<\! h\, \Longleftrightarrow c_{regular} \!<\! c_{singular} = \overline{u}(z_s)\! <\! \overline{u}_0]$: When the phase speed of the singular mode (i.e., the advective speed of the passive wave) is between $\overline{u}_0$ and the regular mode steering level, the passive wave must hinder the counter propagation rate of the regular mode at $z=h$, but not terminate it. Consequently the waves are anti-phased but ${\hat{Q}(z_s)/\hat{Q}(h)} = [G(h,h)-
(h-z_s)]/ G(h,z_s)< 1$, so that the active wave has a greater amplitude of vorticity perturbation than the passive.  This configuration is illustrated in Fig. \ref{fig:4cases}(b).

\item \emph{Passive wave is located below the regular mode steering level} $[-b\!<\! z_s\! <\! z_{sl} =h-G(h,h)\, \Longleftrightarrow  c_{singular} = \overline{u}(z_s) \!<\!  c_{regular}]$: As shown in Fig. \ref{fig:4cases}(d),this is the only case where the two waves are in phase, as the passive wave helps the active one to propagate counter to its local mean flow $\overline{u}_0$. { In this case the vorticity amplitude ratio of the passive and active waves, ${\hat{Q}(z_s)/\hat{Q}(h)} = [(h-z_s)-G(h,h)]/ G(h,z_s)$, can take any (positive) value depending on their relative positions (see also Fig.~\ref{fig:eigen_ratio}).}
\end{enumerate}   



The amplitude ratio of the active and passive waves, $\hat{Q}(z_s)/\hat{Q}(h)$, in the singular modes is plotted in Fig.~\ref{fig:eigen_ratio}, as a function of $z_s/h$.
It is straightforward to understand now why the singular modes are vulnerable to the {numerical grid resolution}.
Since each mode has a vorticity singularity at $z_s$, resulting from discontinuity in $w_{,z}$  at that location, it will remain unresolved if $z_s$ is not sampled by the {grid}. In contrast, the regular mode has no such discontinuity, and should be more robust to differences in the numerical {resolution}.

\section{Non-modal instability: Regular and singular mode interactions} \label{sec:non-modal}

We close the analysis by noting that when $z_s  =z_{sl}$, i.e. when $c_{singular} =  c_{regular}$, we can obtain a non-modal self-sustained wave growth when $\Delta\epsilon(h,z_{sl}) = \pi/2$. Substituting Eq. (\ref{singular_q}) in  Eq. (\ref{instant_gr}) for this phase difference yields
\begin{equation} \label{linear_gr}
\hat{Q}(h,t)= \hat{Q}(h,0) + k {\overline{u}_0 \over h}\hat{Q}(z_{sl})G(h,z_{sl})t,
\end{equation}
where $\hat{Q}(z_{sl})$ remains constant. As illustrated in Fig. \ref{fig:4cases}(c), when $z_s$ is located at the steering level of the regular mode, the two waves are moving in concert without affecting each other's propagation rate. Nonetheless, since the cross-stream velocity field induced by the passive wave is in phase with the vorticity of the active one, it keeps increasing the latter's amplitude by advecting the mean vorticity at $z=h$. We note that since the growth mechanism is only one-way and not mutual (in the sense that the passive wave cannot grow in the absence of a mean vorticity gradient at $z_{sl}$, only the active wave can grow), the growth of $\hat{Q}(h)$ is only linear and not exponential as in modal instability. Thus, this non-modal growth mechanism cannot be detected from the standard perturbation normal mode eigen-analysis of the linearized system. A similar linear growth mechanism has been analyzed by \citet{bishop2000} for a stable baroclinic shear setup. Assuming for simplicity $\hat{Q}(h,0)=0$, the general streamfunction evolution can then be obtained as follows:
{
\begin{equation} \label{linear_psi}
\psi = \mathrm{Re}\left[-\left\{G(z,z_{sl})+\ii k {\overline{u}_0 \over h}G(h,z_{sl})G(z,h)t \right\}\hat{Q}(z_{sl}) 
\mathrm{e}^{\ii(kx- \omega t)}\right],
\end{equation}}
so that after $t \gg G(z,z_{sl})/[k( {\overline{u}_0/ h})G(h,z_{sl})G(z,h)]$, the first term within brackets in the RHS  can be dropped. 
Note the wave's frequency is $\omega= k\overline{u}(z_{sl}) = k\overline{u}_0[1 - {G(h,h)/h}]$.

\begin{figure}
\centering
\includegraphics[width=0.9\linewidth,keepaspectratio]{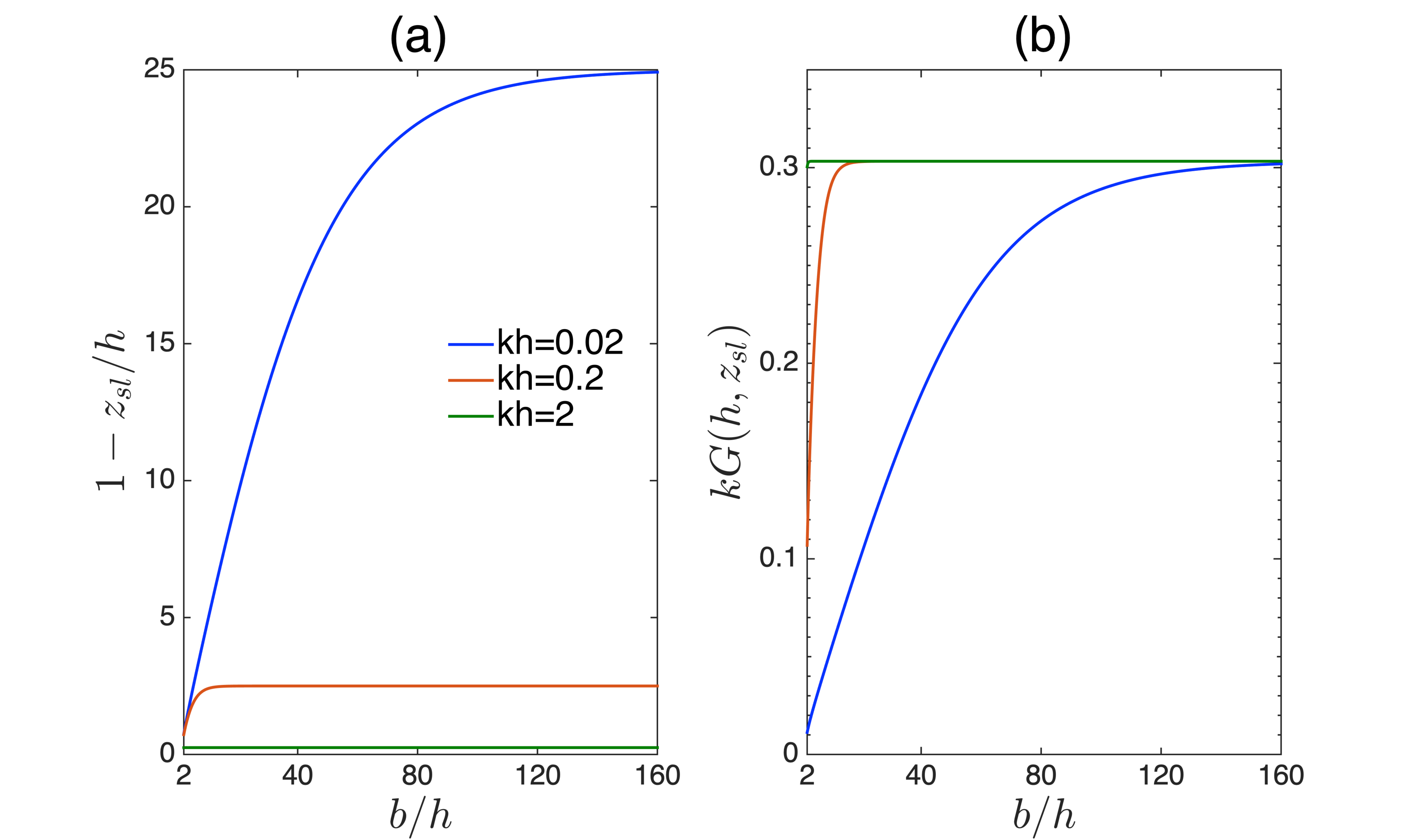}
\caption{{ Effect of domain size ($b/h$) on the non-modal growth: (a) Distance of steering level from active wave, and (b) the linear growth factor that is dependent on the domain size.} }
\label{fig:bound_eff}
\end{figure}

{The boundaries located at $z=\pm b$ impacts the non-modal growth mechanism in a non-trivial way. As $b/h$ increases (boundaries move away farther), the  amplitude of the cross-stream velocity, and hence the intrinsic phase speed of the active wave increases. This causes the active wave to counter-propagate faster, which in turn causes the steering level to move farther away from the active wave, see  Fig.\, \ref{fig:bound_eff}(a). However this relationship does not continue indefinitely since the farther the boundary moves away, its effect on the intrinsic phase speed also decreases. Hence the steering level asymptotes to a constant value (corresponding to $b/h\rightarrow\infty$). As expected, the longer waves are more strongly affected by the boundary effects than the shorter waves. The linear growth rate factor $kG(h,z_{sl})$, which stems from the cross-stream velocity, directly depends on $b/h$. The variation of $kG(h,z_{sl})$ with $b/h$ is  plotted in Fig.\, \ref{fig:bound_eff}(b).
Despite the steering level distance from the active wave ($1-z_{sl}/h$) increases with $b/h$, which in turn is expected to reduce the linear growth rate, the growth rate still increases with $b/h$. This is because the cross-stream velocity from the $z=z_{sl}$ to $z=h$ decays with $b/h$ at a rate slower than the rate with which $1-z_{sl}/h$ increases with $b/h$.  As $b/h$ continues to increase, the boundary effects wane off and the linear growth rate asymptotes to a constant value.}

{
An interesting comparison can be drawn between the modal instability in the Rayleigh's shear profile and the non-modal instability of the profile in question. While in Rayleigh's shear profile, moving boundaries closer stabilizes the flow \cite{heifetz2009canonical}, the profile in question is always unstable to  short waves. For example, Fig.\, \ref{fig:bound_eff}(b) shows that for $b/h=2$, the wavenumber $kh=2$ still grows at a rate $\approx 0.3$.  }
 
In the 
limit of $b/h\rightarrow\infty$, we obtain 
$G(h,h) = 1/2k$ so that $z_{sl} = h-1/2k$ and $\omega= \overline{u}_0(k-1/2h)$ with the constant group velocity   $c_g = {d\omega/ dk} = \overline{u}_0$.
Thus for open boundaries, a linearly growing wavepacket composed of Fourier components of the form of 
{
\begin{equation} \label{linear_psi}
\psi = \mathrm{Re}\left[-\ii{\overline{u}_0 t \over 4h}\mathrm{e}^{-k|h-z_{sl}|}\mathrm{e}^{-k|h-z|} \hat{Q}(z_{sl}) 
\mathrm{e}^{\ii(kx- \omega t)}\right],
\end{equation}}

\noindent will be moving with the speed of the mean flow where the mean vorticity is discontinuous. Since $\hat{Q}(z_{sl})$ is a function of $z_{sl}$ and $z_{sl}$ is a function of $k$, each component of this wavepacket is resonant with a passive wave located on a different level within the shear layer.

\section{Conclusions} \label{sec:conc}


We have demonstrated how wave interaction theory (WIT) can explain the structure of both the neutral regular modes and the continuous spectrum of singular modes in a stable shear layer { (i.e. stable to normal mode perturbations)} in the absence of an inflection point. The regular mode is essentially a shear Rossby/vorticity wave propagating by advecting the local mean vorticity. The singular mode results from an interaction between this ``active" Rossby wave and a singular non-isovortical $\delta$-function vorticity wave that is ``passively" advected by the local mean flow. These two remote waves form a neutral mode by a one-way action-at-a-distance -- the passive singular vorticity wave induces a far field cross-stream velocity that helps or hinders the active Rossby wave to propagate in concert with the former. Since the non-isovortical vorticity perturbation is discontinuous in the cross-stream derivative of the cross-stream velocity (i.e.\, $w_{,z}$), it can be easily missed by a numerical scheme that samples the layers in the $z$ direction. In contrast, the regular mode is robust as it stems from a small but non-zero finite width region whose vorticity perturbation structure results from undulating the mean vorticity isolines.  

Furthermore, when the non-isovortical wave is located at the steering level of the Rossby wave, the two waves can propagate in concert in a quadrature phase-locked setup, where the former is linearly amplifying the latter. For a wave packet in an infinite domain, this forms a linearly growing wave packet propagating with the speed of the local mean flow of the regular mode. Thus WIT provides a natural framework for the understanding of the singular continuous spectrum, its vertical structure, and the interactions leading to algebraic non-modal instabilities.

-----------------------------------------------------------------

Data sharing is not applicable to this article as no new data were created or analyzed in this study.
\section*{Acknowledgements}
The authors thank the two anonymous reviewers for constructive comments and suggestions. A.G. thanks Alexander von Humboldt foundation for funding support. 
\bibliography{smooth_profiles}

\providecommand{\noopsort}[1]{}\providecommand{\singleletter}[1]{#1}%
\begin{thebibliography}{25}%
\makeatletter
\providecommand \@ifxundefined [1]{%
 \@ifx{#1\undefined}
}%
\providecommand \@ifnum [1]{%
 \ifnum #1\expandafter \@firstoftwo
 \else \expandafter \@secondoftwo
 \fi
}%
\providecommand \@ifx [1]{%
 \ifx #1\expandafter \@firstoftwo
 \else \expandafter \@secondoftwo
 \fi
}%
\providecommand \natexlab [1]{#1}%
\providecommand \enquote  [1]{``#1''}%
\providecommand \bibnamefont  [1]{#1}%
\providecommand \bibfnamefont [1]{#1}%
\providecommand \citenamefont [1]{#1}%
\providecommand \href@noop [0]{\@secondoftwo}%
\providecommand \href [0]{\begingroup \@sanitize@url \@href}%
\providecommand \@href[1]{\@@startlink{#1}\@@href}%
\providecommand \@@href[1]{\endgroup#1\@@endlink}%
\providecommand \@sanitize@url [0]{\catcode `\\12\catcode `\$12\catcode
  `\&12\catcode `\#12\catcode `\^12\catcode `\_12\catcode `\%12\relax}%
\providecommand \@@startlink[1]{}%
\providecommand \@@endlink[0]{}%
\providecommand \url  [0]{\begingroup\@sanitize@url \@url }%
\providecommand \@url [1]{\endgroup\@href {#1}{\urlprefix }}%
\providecommand \urlprefix  [0]{URL }%
\providecommand \Eprint [0]{\href }%
\providecommand \doibase [0]{http://dx.doi.org/}%
\providecommand \selectlanguage [0]{\@gobble}%
\providecommand \bibinfo  [0]{\@secondoftwo}%
\providecommand \bibfield  [0]{\@secondoftwo}%
\providecommand \translation [1]{[#1]}%
\providecommand \BibitemOpen [0]{}%
\providecommand \bibitemStop [0]{}%
\providecommand \bibitemNoStop [0]{.\EOS\space}%
\providecommand \EOS [0]{\spacefactor3000\relax}%
\providecommand \BibitemShut  [1]{\csname bibitem#1\endcsname}%
\let\auto@bib@innerbib\@empty
\bibitem [{\citenamefont {Helmholtz}(1868)}]{professor1868xliii}%
  \BibitemOpen
  \bibfield  {author} {\bibinfo {author} {\bibfnamefont {H.}~\bibnamefont
  {Helmholtz}},\ }\bibfield  {title} {\enquote {\bibinfo {title} {{XLIII}. {O}n
  discontinuous movements of fluids},}\ }\href@noop {} {\bibfield  {journal}
  {\bibinfo  {journal} {Philos. Mag.}\ }\textbf {\bibinfo {volume} {36}},\
  \bibinfo {pages} {337--346} (\bibinfo {year} {1868})}\BibitemShut {NoStop}%
\bibitem [{\citenamefont {Thomson}(1871)}]{thomson1871xlvi}%
  \BibitemOpen
  \bibfield  {author} {\bibinfo {author} {\bibfnamefont {W.}~\bibnamefont
  {Thomson}},\ }\bibfield  {title} {\enquote {\bibinfo {title} {{XLVI}.
  {H}ydrokinetic solutions and observations},}\ }\href@noop {} {\bibfield
  {journal} {\bibinfo  {journal} {Philos. Mag.}\ }\textbf {\bibinfo {volume}
  {42}},\ \bibinfo {pages} {362--377} (\bibinfo {year} {1871})}\BibitemShut
  {NoStop}%
\bibitem [{\citenamefont {Rayleigh}(1880)}]{rayl1880}%
  \BibitemOpen
  \bibfield  {author} {\bibinfo {author} {\bibfnamefont {J.~W.~S.}\
  \bibnamefont {Rayleigh}},\ }\bibfield  {title} {\enquote {\bibinfo {title}
  {On the stability, or instability, of certain fluid motions},}\ }\href@noop
  {} {\bibfield  {journal} {\bibinfo  {journal} {Proc. Lond. Math. Soc.}\
  }\textbf {\bibinfo {volume} {12}},\ \bibinfo {pages} {57--70} (\bibinfo
  {year} {1880})}\BibitemShut {NoStop}%
\bibitem [{\citenamefont {Orr}(1907)}]{orr1907}%
  \BibitemOpen
  \bibfield  {author} {\bibinfo {author} {\bibfnamefont {W.~M.~F.}\
  \bibnamefont {Orr}},\ }\bibfield  {title} {\enquote {\bibinfo {title}
  {Stability or instability of the steady motions of a perfect liquid and of a
  viscous liquid},}\ }\href@noop {} {\bibfield  {journal} {\bibinfo  {journal}
  {Proc. Roy. Irish Acad.}\ }\textbf {\bibinfo {volume} {A}},\ \bibinfo {pages}
  {9--138} (\bibinfo {year} {1907})}\BibitemShut {NoStop}%
\bibitem [{\citenamefont {Sommerfeld}(1909)}]{sommerfeld1909}%
  \BibitemOpen
  \bibfield  {author} {\bibinfo {author} {\bibfnamefont {A.}~\bibnamefont
  {Sommerfeld}},\ }\bibfield  {title} {\enquote {\bibinfo {title} {A
  contribution to the hydrodynamical explanation of turbulent fluid motions},}\
  }\href@noop {} {\bibfield  {journal} {\bibinfo  {journal} {Proc. 4-th
  Internat. Congress of Mathematicians Rome}\ ,\ \bibinfo {pages} {116--124}}
  (\bibinfo {year} {1909})}\BibitemShut {NoStop}%
\bibitem [{\citenamefont {Fj{\o}rtoft}(1953)}]{fjortoft1953}%
  \BibitemOpen
  \bibfield  {author} {\bibinfo {author} {\bibfnamefont {R.}~\bibnamefont
  {Fj{\o}rtoft}},\ }\bibfield  {title} {\enquote {\bibinfo {title} {On the
  changes in the spectral distribution of kinetic energy for twodimensional,
  nondivergent flow},}\ }\href@noop {} {\bibfield  {journal} {\bibinfo
  {journal} {Tellus}\ }\textbf {\bibinfo {volume} {5}},\ \bibinfo {pages}
  {225--230} (\bibinfo {year} {1953})}\BibitemShut {NoStop}%
\bibitem [{\citenamefont {Drazin}\ and\ \citenamefont
  {Reid}(2004)}]{drazin2004hydrodynamic}%
  \BibitemOpen
  \bibfield  {author} {\bibinfo {author} {\bibfnamefont {P.~G.}\ \bibnamefont
  {Drazin}}\ and\ \bibinfo {author} {\bibfnamefont {W.~H.}\ \bibnamefont
  {Reid}},\ }\href@noop {} {\emph {\bibinfo {title} {Hydrodynamic stability}}}\
  (\bibinfo  {publisher} {Cambridge university press},\ \bibinfo {year}
  {2004})\BibitemShut {NoStop}%
\bibitem [{\citenamefont {Schmid}\ and\ \citenamefont
  {Henningson}(2001)}]{schmid2001stability}%
  \BibitemOpen
  \bibfield  {author} {\bibinfo {author} {\bibfnamefont {P.}~\bibnamefont
  {Schmid}}\ and\ \bibinfo {author} {\bibfnamefont {D.}~\bibnamefont
  {Henningson}},\ }\href@noop {} {\emph {\bibinfo {title} {Stability and
  Transition in Shear Flows}}},\ Vol.\ \bibinfo {volume} {142}\ (\bibinfo
  {publisher} {Springer Verlag},\ \bibinfo {year} {2001})\BibitemShut {NoStop}%
\bibitem [{\citenamefont {Bender}\ and\ \citenamefont
  {Orszag}(2013)}]{bender2013advanced}%
  \BibitemOpen
  \bibfield  {author} {\bibinfo {author} {\bibfnamefont {C.~M.}\ \bibnamefont
  {Bender}}\ and\ \bibinfo {author} {\bibfnamefont {S.~A.}\ \bibnamefont
  {Orszag}},\ }\href@noop {} {\emph {\bibinfo {title} {Advanced mathematical
  methods for scientists and engineers I: Asymptotic methods and perturbation
  theory}}}\ (\bibinfo  {publisher} {Springer Science \& Business Media},\
  \bibinfo {year} {2013})\BibitemShut {NoStop}%
\bibitem [{\citenamefont {Balmforth}\ and\ \citenamefont
  {Morrison}(1995)}]{balmforth1995normal}%
  \BibitemOpen
  \bibfield  {author} {\bibinfo {author} {\bibfnamefont {N.}~\bibnamefont
  {Balmforth}}\ and\ \bibinfo {author} {\bibfnamefont {P.~J.}\ \bibnamefont
  {Morrison}},\ }\bibfield  {title} {\enquote {\bibinfo {title} {Normal modes
  and continuous spectra},}\ }\href@noop {} {\bibfield  {journal} {\bibinfo
  {journal} {Ann. N. Y. Acad. Sci.}\ }\textbf {\bibinfo {volume} {773}},\
  \bibinfo {pages} {80--94} (\bibinfo {year} {1995})}\BibitemShut {NoStop}%
\bibitem [{\citenamefont {Drazin}(2002)}]{drazin2002introduction}%
  \BibitemOpen
  \bibfield  {author} {\bibinfo {author} {\bibfnamefont {P.~G.}\ \bibnamefont
  {Drazin}},\ }\href@noop {} {\emph {\bibinfo {title} {Introduction to
  Hydrodynamic Stability}}},\ Vol.~\bibinfo {volume} {32}\ (\bibinfo
  {publisher} {Cambridge university press},\ \bibinfo {year}
  {2002})\BibitemShut {NoStop}%
\bibitem [{\citenamefont {Carpenter}\ and\ \citenamefont
  {Guha}(2019)}]{carp19}%
  \BibitemOpen
  \bibfield  {author} {\bibinfo {author} {\bibfnamefont {J.~R.}\ \bibnamefont
  {Carpenter}}\ and\ \bibinfo {author} {\bibfnamefont {A.}~\bibnamefont
  {Guha}},\ }\bibfield  {title} {\enquote {\bibinfo {title} {Instability of a
  smooth shear layer through wave interactions},}\ }\href@noop {} {\bibfield
  {journal} {\bibinfo  {journal} {Phys. Fluids}\ }\textbf {\bibinfo {volume}
  {31}},\ \bibinfo {pages} {081701} (\bibinfo {year} {2019})}\BibitemShut
  {NoStop}%
\bibitem [{\citenamefont {Iga}(2013)}]{iga2013}%
  \BibitemOpen
  \bibfield  {author} {\bibinfo {author} {\bibfnamefont {K.}~\bibnamefont
  {Iga}},\ }\bibfield  {title} {\enquote {\bibinfo {title} {Shear instability
  as a resonance between neutral waves hidden in a shear flow},}\ }\href@noop
  {} {\bibfield  {journal} {\bibinfo  {journal} {J. Fluid Mech.}\ }\textbf
  {\bibinfo {volume} {715}},\ \bibinfo {pages} {452--476} (\bibinfo {year}
  {2013})}\BibitemShut {NoStop}%
\bibitem [{\citenamefont {Hoskins}, \citenamefont {McIntyre},\ and\
  \citenamefont {Robertson}(1985)}]{hosk1985}%
  \BibitemOpen
  \bibfield  {author} {\bibinfo {author} {\bibfnamefont {B.~J.}\ \bibnamefont
  {Hoskins}}, \bibinfo {author} {\bibfnamefont {M.~E.}\ \bibnamefont
  {McIntyre}}, \ and\ \bibinfo {author} {\bibfnamefont {A.~W.}\ \bibnamefont
  {Robertson}},\ }\bibfield  {title} {\enquote {\bibinfo {title} {On the use
  and significance of isentropic potential vorticity maps},}\ }\href@noop {}
  {\bibfield  {journal} {\bibinfo  {journal} {Q. J. Roy. Meteor. Soc.}\
  }\textbf {\bibinfo {volume} {111}},\ \bibinfo {pages} {877--946} (\bibinfo
  {year} {1985})}\BibitemShut {NoStop}%
\bibitem [{\citenamefont {Baines}\ and\ \citenamefont
  {Mitsudera}(1994)}]{bain1994}%
  \BibitemOpen
  \bibfield  {author} {\bibinfo {author} {\bibfnamefont {P.}~\bibnamefont
  {Baines}}\ and\ \bibinfo {author} {\bibfnamefont {H.}~\bibnamefont
  {Mitsudera}},\ }\bibfield  {title} {\enquote {\bibinfo {title} {On the
  mechanism of shear flow instabilities},}\ }\href@noop {} {\bibfield
  {journal} {\bibinfo  {journal} {J. Fluid Mech.}\ }\textbf {\bibinfo {volume}
  {276}},\ \bibinfo {pages} {327--342} (\bibinfo {year} {1994})}\BibitemShut
  {NoStop}%
\bibitem [{\citenamefont {Smyth}\ and\ \citenamefont {Carpenter}(2019)}]{book}%
  \BibitemOpen
  \bibfield  {author} {\bibinfo {author} {\bibfnamefont {W.~D.}\ \bibnamefont
  {Smyth}}\ and\ \bibinfo {author} {\bibfnamefont {J.~R.}\ \bibnamefont
  {Carpenter}},\ }\href@noop {} {\emph {\bibinfo {title} {Instability in
  Geophysical Flows}}}\ (\bibinfo  {publisher} {Cambridge University Press},\
  \bibinfo {year} {2019})\BibitemShut {NoStop}%
\bibitem [{\citenamefont {Carpenter}\ \emph {et~al.}(2013)\citenamefont
  {Carpenter}, \citenamefont {Tedford}, \citenamefont {Heifetz},\ and\
  \citenamefont {Lawrence}}]{carp2013}%
  \BibitemOpen
  \bibfield  {author} {\bibinfo {author} {\bibfnamefont {J.~R.}\ \bibnamefont
  {Carpenter}}, \bibinfo {author} {\bibfnamefont {E.~W.}\ \bibnamefont
  {Tedford}}, \bibinfo {author} {\bibfnamefont {E.}~\bibnamefont {Heifetz}}, \
  and\ \bibinfo {author} {\bibfnamefont {G.~A.}\ \bibnamefont {Lawrence}},\
  }\bibfield  {title} {\enquote {\bibinfo {title} {Instability of stratified
  shear flow: review of a physical mechanism based on interacting waves},}\
  }\href@noop {} {\bibfield  {journal} {\bibinfo  {journal} {Appl. Mech. Rev.}\
  }\textbf {\bibinfo {volume} {64}},\ \bibinfo {pages} {060801} (\bibinfo
  {year} {2013})}\BibitemShut {NoStop}%
\bibitem [{\citenamefont {Holmboe}(1962)}]{holm1962}%
  \BibitemOpen
  \bibfield  {author} {\bibinfo {author} {\bibfnamefont {J.}~\bibnamefont
  {Holmboe}},\ }\bibfield  {title} {\enquote {\bibinfo {title} {On the behavior
  of symmetric waves in stratified shear layers},}\ }\href@noop {} {\bibfield
  {journal} {\bibinfo  {journal} {Geofys. Publ.}\ }\textbf {\bibinfo {volume}
  {24}},\ \bibinfo {pages} {67--112} (\bibinfo {year} {1962})}\BibitemShut
  {NoStop}%
\bibitem [{\citenamefont {Bretherton}(1966)}]{bretherton}%
  \BibitemOpen
  \bibfield  {author} {\bibinfo {author} {\bibfnamefont {F.~P.}\ \bibnamefont
  {Bretherton}},\ }\bibfield  {title} {\enquote {\bibinfo {title} {Baroclinic
  instability and the short wavelength cut-off in terms of potential
  vorticity},}\ }\href@noop {} {\bibfield  {journal} {\bibinfo  {journal} {Q.
  J. R. Meteorol. Soc.}\ }\textbf {\bibinfo {volume} {92}},\ \bibinfo {pages}
  {335--345} (\bibinfo {year} {1966})}\BibitemShut {NoStop}%
\bibitem [{\citenamefont {Heifetz}\ and\ \citenamefont
  {Methven}(2005)}]{heif2005}%
  \BibitemOpen
  \bibfield  {author} {\bibinfo {author} {\bibfnamefont {E.}~\bibnamefont
  {Heifetz}}\ and\ \bibinfo {author} {\bibfnamefont {J.}~\bibnamefont
  {Methven}},\ }\bibfield  {title} {\enquote {\bibinfo {title} {Relating
  optimal growth to counterpropagating {R}ossby waves in shear instability},}\
  }\href@noop {} {\bibfield  {journal} {\bibinfo  {journal} {Phys. Fluids}\
  }\textbf {\bibinfo {volume} {17}},\ \bibinfo {eid} {064107} (\bibinfo {year}
  {2005})}\BibitemShut {NoStop}%
\bibitem [{\citenamefont {Redekopp}(2001)}]{rede2001}%
  \BibitemOpen
  \bibfield  {author} {\bibinfo {author} {\bibfnamefont {L.}~\bibnamefont
  {Redekopp}},\ }\bibfield  {title} {\enquote {\bibinfo {title} {Elements of
  instability theory for environmental flows},}\ }in\ \href@noop {} {\emph
  {\bibinfo {booktitle} {Environmental Stratified Flows}}}\ (\bibinfo
  {publisher} {Kluwer, Boston},\ \bibinfo {year} {2001})\BibitemShut {NoStop}%
\bibitem [{\citenamefont {Heifetz}\ and\ \citenamefont
  {Guha}(2019)}]{heif2019normal}%
  \BibitemOpen
  \bibfield  {author} {\bibinfo {author} {\bibfnamefont {E.}~\bibnamefont
  {Heifetz}}\ and\ \bibinfo {author} {\bibfnamefont {A.}~\bibnamefont {Guha}},\
  }\bibfield  {title} {\enquote {\bibinfo {title} {Normal form of
  synchronization and resonance between vorticity waves in shear flow
  instability},}\ }\href@noop {} {\bibfield  {journal} {\bibinfo  {journal}
  {Phys. Rev. E}\ }\textbf {\bibinfo {volume} {100}},\ \bibinfo {pages}
  {043105} (\bibinfo {year} {2019})}\BibitemShut {NoStop}%
\bibitem [{\citenamefont {Baines}, \citenamefont {Majumdar},\ and\
  \citenamefont {Mitsudera}(1996)}]{bain1996}%
  \BibitemOpen
  \bibfield  {author} {\bibinfo {author} {\bibfnamefont {P.~G.}\ \bibnamefont
  {Baines}}, \bibinfo {author} {\bibfnamefont {S.}~\bibnamefont {Majumdar}}, \
  and\ \bibinfo {author} {\bibfnamefont {H.}~\bibnamefont {Mitsudera}},\
  }\bibfield  {title} {\enquote {\bibinfo {title} {The mechanics of the
  {T}ollmein-{S}chlichting wave},}\ }\href@noop {} {\bibfield  {journal}
  {\bibinfo  {journal} {J. Fluid Mech.}\ }\textbf {\bibinfo {volume} {312}},\
  \bibinfo {pages} {107--124} (\bibinfo {year} {1996})}\BibitemShut {NoStop}%
\bibitem [{\citenamefont {Bishop}\ and\ \citenamefont
  {Heifetz}(2000)}]{bishop2000}%
  \BibitemOpen
  \bibfield  {author} {\bibinfo {author} {\bibfnamefont {C.~H.}\ \bibnamefont
  {Bishop}}\ and\ \bibinfo {author} {\bibfnamefont {E.}~\bibnamefont
  {Heifetz}},\ }\bibfield  {title} {\enquote {\bibinfo {title} {Apparent
  absolute instability and the continuous spectrum},}\ }\href@noop {}
  {\bibfield  {journal} {\bibinfo  {journal} {J. Atmos. Sci.}\ }\textbf
  {\bibinfo {volume} {57}},\ \bibinfo {pages} {3592--3608} (\bibinfo {year}
  {2000})}\BibitemShut {NoStop}%
\bibitem [{\citenamefont {Heifetz}, \citenamefont {Harnik},\ and\ \citenamefont
  {Tamarin}(2009)}]{heifetz2009canonical}%
  \BibitemOpen
  \bibfield  {author} {\bibinfo {author} {\bibfnamefont {E.}~\bibnamefont
  {Heifetz}}, \bibinfo {author} {\bibfnamefont {N.}~\bibnamefont {Harnik}}, \
  and\ \bibinfo {author} {\bibfnamefont {T.}~\bibnamefont {Tamarin}},\
  }\bibfield  {title} {\enquote {\bibinfo {title} {Canonical hamiltonian
  representation of pseudoenergy in shear flows using counter-propagating
  rossby waves},}\ }\href@noop {} {\bibfield  {journal} {\bibinfo  {journal}
  {Q. J. Roy. Meteor. Soc.}\ }\textbf {\bibinfo {volume} {135}},\ \bibinfo
  {pages} {2161--2167} (\bibinfo {year} {2009})}\BibitemShut {NoStop}%
\end{thebibliography}%

\end{document}